\documentclass[10pt,a4paper,twoside,twocolumn]{article}

\usepackage[T1]{fontenc}
\usepackage{ae}
\usepackage[latin1]{inputenc}
\usepackage{graphicx}
\usepackage{color}
\usepackage{hyperref}

\newcommand{\ket}[1]{|{#1}\rangle}
\newcommand{\bra}[1]{\langle{#1}|}
\newcommand{\av}[1]{\langle{#1}\rangle}
\newcommand{\dd}{\,\mathrm{d}}

\newcommand{\im}{\,\mathcal{I}\!\mathit{m}\,}
\newcommand{\op}[1]{\hat{#1}}
\newcommand{\tr}{\mathop{\mathrm{tr}}}

\newcommand{\mat}[2]{\left(\!\!\begin{array}{#1} #2 \end{array}\!\!\right)}

\renewcommand{\tanh}{\mathop{\mathrm{th}}}

\newcommand{\Argth}{\mathop{\mathrm{Argth}}}

\begin{document}

\title{
	Loss of quantum coherence in a system coupled to a 
	zero-temperature environment
	}

\author{
	A.~Ratchov
	F.~Faure and 
	F.~W.~J.~Hekking
	\\
	\small{ Laboratoire de physique et de
		modélisation des milieux condensées,
		université Joseph Fourier \& \textsc{cnrs} }
	\\
	\small{ 25,
		av.~des~Martyrs, \textsc{bp}~166,
		38042 \textsc{cedex},
		France
		}
	}

\date{December 22, 2003}

\maketitle

\noindent 
PACS numbers~: 03.65.Yz

\begin{abstract}
	We discuss the influence of a zero-temperature environment on a
	coherent quantum system. First, we calculate the reduced density operator
	of the system in the framework of the well-known, exactly
	solvable model of an oscillator coupled to a bath of harmonic
	oscillators. Then, we propose the sketch of an
	Aharonov--Bohm-like interferometer showing, through
	interference measurements, the decrease of the coherence length
	of the system due to the interaction with the environment,
	even in the zero temperature limit.
	\end{abstract}


\section{Introduction}


	The effects of the interaction between a quantum system and its
	environment have been studied since the early days of quantum
	mechanics. For instance, in quantum measurement theory, a study
	of the role of the environment helps to understand the
	transition between the quantum and the classical world
	(see~\cite{deco}). Another example is quantum electrodynamics
	(see \cite{miloni}), where to some extent one can consider the
	electromagnetic field as the environment, influencing a charged
	particle (the quantum system).

	Clearly, these issues are of key importance in mesoscopic
	physics. Since the discovery of mesoscopic phenomena in
	solids~\cite{imry}, it is well-known that the transport
	properties of small metallic systems at low temperatures are
	strongly influenced by interference of electronic waves.
	Examples are the weak localisation correction to conductivity
	or the universal sample-to-sample fluctuations of the
	conductance. On the other hand, the quantum behaviour of
	``free'' electrons in mesoscopic systems is affected by their
	interaction with the environment, which for example consists of
	other electrons, phonons, photons, or scatterers. Which
	environment dominates the destruction of interference
	phenomena, an effect sometimes referred to as decoherence,
	depends in general on the temperature. For instance, the
	temperature dependence of the weak-localisation correction to
	the conductivity reveals that in metals electron-electron
	interactions dominate over the phonon contribution to
	decoherence at the lowest temperatures.


	In this connection, the question as to what happens to
	interference phenomena at zero temperature has been hotly
	debated over the past few years. This debate was initiated by
	temperature-dependent weak localisation
	measurements~\cite{mohanty}, reporting on a residual
	decoherence in metals at zero temperature, in contradiction
	with theoretical predictions~\cite{altshuler}. The subsequent
	theoretical debate~\cite{golubev,aleiner,vondelft} mainly
	focused on zero-temperature decoherence induced by Coulomb
	interactions in disordered electron systems, but as a spin-off
	has led to the more general question ``Can a
	zero-temperature environment induce decoherence?''. Recently,
	this issue was discussed in \cite{buttiker,jordan} in
	the framework of a well-known, simple model~: a harmonic
	oscillator (the ``particle'') coupled to a chain of harmonic
	oscillators (the ``environment''). It was shown that the
	particle exchanges energy with the environment, even at zero
	temperature. The effect of these energy fluctuations cannot be
	simply captured through a renormalisation of the particle's
	parameters, but will give rise to a ground-state with
	non-trivial dynamics. This can have important consequences on
	thermodynamic properties of measured systems; an example is the
	suppression of the zero-temperature persistent normal or 
	super-current in mesoscopic rings~\cite{hekking,cedraschi}.


	In the present paper, we are interested in the influence of an
	environment at low temperature on the behaviour of a mesoscopic
	system. We are in particular interested in the effects of the
	environment the interference phenomena, in the limit when the
	coupling energy between a small system and the environment is
	larger than the thermal energy. 
	In the following, we consider a simple, exactly solvable model
	of a particle coupled to the environment. We regard the
	particle as part of a larger system of a particle coupled
	to a heat bath and calculate the exact reduced density operator
	of the particle.  Then  we propose the sketch of a device
	showing the saturation of the coherence length of the particle
	with decreasing temperature. The coherence length is investigated in
	the device by an Aharonov-Bohm interference measurement. 

		\begin{figure}
		\begin{center}
		\mbox{\input{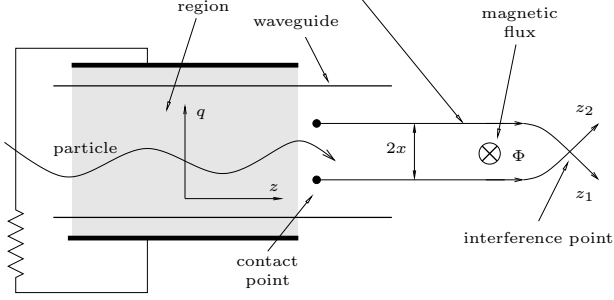}}
		\end{center}
		\caption{A particle travelling longitudinally in the
		waveguide, while transversely coupled to the environment
		is probed by two leads.}
		\label{fig:guide}
		\end{figure}

\begin{figure}[t]
		\begin{center}
		\input{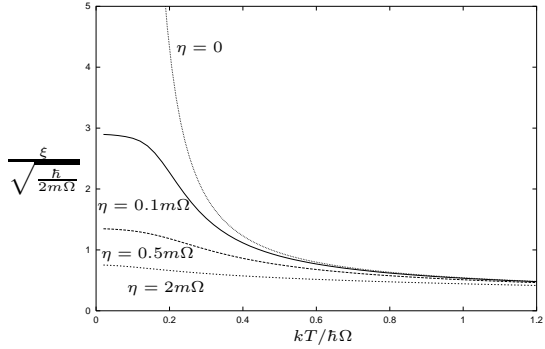}
		\end{center}
		\caption{Coherence length $\xi$ as function of the temperature
			$T$ for several strengths $\eta$ of the coupling of the
			particle and the environment. The main effect
			is the saturation of $\xi(T)$ at finite value,
			when $T\rightarrow 0$.}
		\label{fig:leff}
		\end{figure}

	The model, largely inspired by \cite{buttiker,cl,ford}, is a
	single particle moving longitudinally in a perfect waveguide while
	being transversely coupled to a continuous set of independent
	oscillators (the ``environment''), see fig.~\ref{fig:guide}.
	We probe the state of the particle in the
	waveguide with two perfect 1--dimensional leads in order to
	observe Aharonov-Bohm interferences; the contrast of the
	interference fringes is related to the transverse
	coherence length of the particle in the waveguide. 
	The model is simple enough to do all computations without
	approximations. We will see that the coupling $\eta$ between the
	particle and the environment destroys the interference
	fringes and thus reduces the coherence length $\xi$ of the
	particle (fig.~\ref{fig:leff}). These effects remain even in
	the zero temperature limit.
		%


        This paper is organised as follows~:
	in section~\ref{sec:matrice}, we consider the effects induced
	on a little system by the environment; in
	section~\ref{sec:interf}, we present the Aharonov-Bohm-like
	interferometer capable of showing that the coherence length of
	the particle saturates in the zero-temperature limit. The
	calculations are worked out in appendix \ref{sec:annexe1} and
	\ref{sec:annexe2}.

\section{A particle coupled to the environment}
\label{sec:matrice}

\subsection{The model}

	In order to get exact results, we will use the simplest
	well-known, non-trivial model for a particle coupled to an
	environment~: a harmonic oscillator coupled to a set of $N$
	independent harmonic oscillators \cite{weiss}.
	Consider the classical hamiltonian of
	the whole system~:
		\begin{eqnarray}
		\label{eq:hamtot}
		H(q,p,\varphi,\pi) &=& 
		\underbrace{\frac{p^2}{2m} +
		\frac{m\Omega^2}{2}q^2}_{\mathrm{particle}} 
		\\
		\nonumber
		&+&
		\underbrace{
		\sum_{i=1}^{N} 
		\left\{
		\frac{\pi_i^2}{2\mu_i} +
		\frac{\mu_i \omega_i^2}{2}
		\left(
		\varphi_i - q
		\right)^2
		\right\}
		}_{\mathrm{environment}}
		\end{eqnarray}
	The first part represents the ``particle'' (of mass $m$,
	frequency $\Omega$, position $q$ and momentum $p$).
	The second term corresponds to a set of $N$ independent
	harmonic oscillators, 
		%
		%
	$\omega_i$ is the frequency of the $i$th oscillator,  $\mu_i$,
	$\pi_i$ and $\varphi_i$ are its mass, momentum and
	position respectively. 
	Parameters $\mu_i$ and $\omega_i$ characterise entirely the
	environment. One can see that in the
	zero-temperature limit,  the energy of the whole system is zero
	and, therefore, the particle is at rest,  \emph{i.e.} is in
	its classical ground state. Later, we will see that this
	statement is no longer correct in quantum mechanics.
	
	We will be interested in the behaviour of the system
	when $N$ is large, particularly in the continuum limit. In
	this case, $\omega$ is a continuous variable and the mass
	distribution $\mu_i$ is a smooth function
	$\mu(\omega)$ defined such that $\mu(\omega)\dd\omega$ is
	the mass of the oscillators with frequency between  $\omega$
	and $\omega+\dd\omega$; the distribution  $\mu(\omega)$
	characterises entirely the environment\footnote{
		often, instead of $\mu(\omega)$ 
		we use the spectral function 
		of the environment \cite{cl,weiss}, defined by~:
			\begin{equation}%
			J(\omega) = \pi\omega^3\mu(\omega)
			\end{equation}%
		}.
	In that case, one can prove \cite{cl,ford} that for
	the particular case $\mu(\omega) = 2\eta/m\omega^2$, the
	hamiltonian leads to the well-known classical equation of
	motion~:
		\begin{equation}
		m\ddot q 
		=
		-
		m \Omega^2 q
		\,
		\underbrace{- \,
		\eta\dot q 
		+ 
		F(t)}_{\mathrm{env.}}
		\label{eq:langevin}
		\end{equation}
	The environment induces a dissipative force $-\eta\dot q$ and a
	fluctuating force $F(t)$. The parameter $\eta/m\Omega$ defines the strength of
	the coupling. For that particular environment, $F(t)$
	is a white noise. For other functions $\mu(\omega)$,
	the environment may induce an equation of motion with non-locality in
	time and coloured noise. In this section, we will consider any
	distribution $\mu(\omega)$.
	
\subsection {Reduced density operator of the particle}

	We consider now the quantum version of the system defined by
	eq.~\ref{eq:hamtot}, \emph{i.e.} a quantum oscillator coupled
	to a quantum environment. We replace the classical variables
	by canonical operators~:
		\begin{eqnarray*}
		q, p, \varphi, \pi, \dots
		&\longrightarrow&
		\op q, \op p, \op \varphi, \op \pi, \dots
		\end{eqnarray*}	
	
	We suppose that the entire system is in equilibrium
	at temperature $T$, its density operator is the following~:
		\begin{equation}%
		\op{\rho} = \frac{1}{Z}e^{-\op{H}/kT}
		\textrm{\quad where \quad}
		Z = \tr \op \rho
		\end{equation}%
		
	Generally speaking for a coupled system, $\op\rho$ cannot
	be represented by a tensor product of a state of the particle
	and a state of the environment. The reduced density operator $\op{\sigma}$ of the particle  is
	defined as the trace over the environment of $\op\rho$~:
		\begin{equation}%
		\op\sigma = \tr_{\mathrm{env.}}{\op\rho}
		\end{equation}%
	$\op\sigma$ completely describes  the state of the particle 
	in the sense that it predicts any measurement made on it.
	As shown in appendix \ref{sec:annexe1}, 
	we can write $\op{\sigma}$ in a canonical form
	with unknown coefficients $\tilde T$ and $\tilde m$~:
		\begin{equation}
		\label{eq:sigma} 
		\op{\sigma} = \frac{1}{\tilde Z} \exp \left(
			- \frac{1}{k\tilde T}
				\op H_{\mathrm{eff}}
				\right)
		\end{equation}
	where $\op H_{\mathrm{eff}}$ is the effective
	hamiltonian defined by~:
		\begin{equation}
		\label{eq:Heff} 
		\op H_{\mathrm{eff}} = 
				\frac{\op{p}^2}{2\tilde m} + 
				\frac{\tilde m \Omega^2}{2}\op{q}^2
		\end{equation}
	$\tilde T$ can
	be interpreted as an effective temperature \emph{a priori}
	different from real temperature $T$ because of the
	coupling  with the environment. Similarly, the effective
	mass $\tilde m$ differs
	from the mass $m$ of the particle. In particular we will see that
	even at zero temperature $T=0$, we may have $\tilde T > 0$.
	Therefore, $\op \sigma$ is not
	a pure state, in particular, the particle cannot 
	be in its ground state. Thus, parameters $\tilde T$ and
	$\tilde m$ necessarily differ from
	the real temperature $T$ and the real mass $m$ of the particle.
	$\tilde Z$ is a normalisation constant that ensures that
	$\tr \op\sigma=1$. We can explicitly calculate $\tilde T$
	and $\tilde m$ as function of the real temperature $T$
	and the mass distribution $\mu(\omega)$ of 
	the environment (see appendix \ref{sec:annexe1}).

	This study remains true for any function $\mu(\omega)$;
	however if there is a large gap in $\mu(\omega)$ around
	$\Omega$, the reduced resolvant (discussed in 
	appendix~\ref{sec:annexe1}) can have a pole on the real axis.
	It corresponds to a discrete oscillating mode, analogue
	of a dressed state in atomic physics. Because of
	the coupling between particle and environment, this mode has
	components on both the particle and the environment. If it is
	initially excited then it will oscillate forever and the
	particle will never reach equilibrium. In a future work
	\cite{future} this issue will be discussed and we will show that if~:
		\begin{equation}
		\tr_{\mathrm{part}} \op\rho(t=0) 
			= \frac{e^{\op H_{\mathrm{env}}/kT}}{Z_{\mathrm{env}}}
		\end{equation}
	and $\mu(\omega)$ has no gap, then~:
		\begin{equation}
		\lim_{t\rightarrow 0}
			\tr_{\mathrm{env}} \op\rho(t) = \op\sigma
		\end{equation}
	In	sec.~\ref{sec:interf}, we will suppose that $\mu(\omega)$ is
	such that the reduced resolvant of the  system has no poles on
	the real axis.

\subsection{An example --- ohmic environment}
	\label{sec:exple1}
\begin{figure}[p]
			\begin{center}
			\mbox{ \input{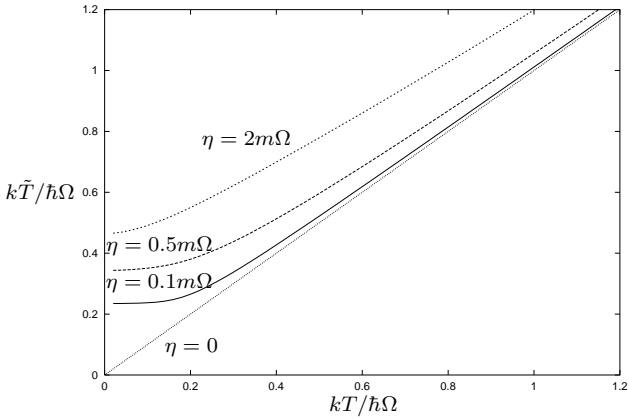} }
			\end{center}
			\caption{
				Effective temperature $\tilde T$ 
				as function of the temperature $T$,
				plotted for different values
				of the coupling $\eta$.
				}
			\label{fig:ttilde}
		\end{figure}

	In order to illustrate the results discussed above, 
	consider the case of an ohmic environment, defined by~:
		\begin{equation}%
		\mu(\omega) = \left\{
			\begin{array}{llc}
			2\eta/\pi\omega^2 &\textrm{if}& \omega<\Omega_c
			\\
			0 &\textrm{otherwise}&
			\end{array}
			\right.
		\end{equation}%
	In this case, for $\Omega_c\gg \Omega$, the classical behaviour of the
	particle is described by
	eq.~\ref{eq:langevin}, see~\cite{buttiker,cl,ford}.  
	$\tilde T(T)$ can be calculated explicitly 
	(see appendix~\ref{sec:annexe1}). It is plotted for different 
	values of the coupling $\eta$ between the particle and the
	environment in fig.~\ref{fig:ttilde}.
	At $T=0$, we find the following limit expression 
	for small $\eta$~:
		\begin{eqnarray}
		k\tilde T(0) &\sim&  
			\frac{\hbar\Omega}{2} \times
			\frac{2}{\ln(2\pi/\gamma)}
			\\
			\nonumber
			&&\textrm{where \quad} 
			\gamma = 
			\frac{\eta}{m\Omega}\ln(\Omega_c/\Omega)
		\end{eqnarray}
	We  can see that, even at $T=0$, the behaviour of the particle
	is very similar to the behaviour of a particle at strictly positive
	temperature~: its reduced density operator is not  a pure state
	as it would be for an isolated particle, but a statistical mixture.
	
	Similarly $\tilde m(T)$ can be calculated explicitly
	(see appendix~\ref{sec:annexe1}). It is plotted for different 
	values of $\eta$ in fig.~\ref{fig:mtilde}.
	At $T=0$, we find the following limit expressions,
	for small $\eta$~:
		\begin{eqnarray}
		\tilde m(0) &\sim&
			m \left( 1 + \gamma/\pi \right)
		\end{eqnarray}
		\begin{figure}[p]
			\begin{center}
			\mbox{
			\input{t_meff.pstex_t}
			}
			\end{center}
			\caption{
				Effective mass $\tilde m$ 
				as function of the temperature $T$,
				plotted for different values
				of the coupling $\eta$.
				}
			\label{fig:mtilde}
		\end{figure}
		\begin{figure}[p]
			\begin{center}
			\mbox{
			\input{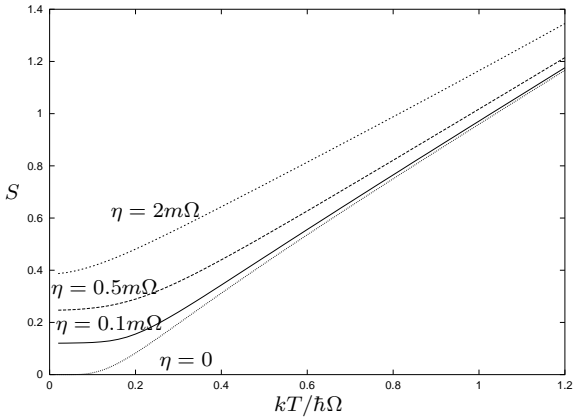}
			}
			\end{center}
			\caption{
				Entropy $S$ of the particle as function
				of the temperature $T$, plotted for different
				values of the coupling $\eta$.
				}			
			\label{fig:t_seff}
		\end{figure}
	It is interesting to evaluate the entropy of the particle, 
	defined by $S=-\tr(\op{\sigma}\ln\op{\sigma})$. 
	In contrast to $\tilde T$ and $\tilde m$, $S$ has
	an intrinsic definition. On fig.~\ref{fig:t_seff},
	we see that there is 
	a residual entropy that does not vanish at $T=0$.	 
	These statements are not really in contradiction with
	the ordinary statistical mechanics (or with the 3-rd
	principle of thermodynamics); indeed in statistical mechanics
	one neglects the coupling energy between the
	particle and the reservoir. That approximation is good at high
	temperature when $T\gg \tilde T(0)$.

\section{Interferences in a device}
\label{sec:interf}
		
	In order to illustrate the physical meaning of these
	results, in the following section we propose the
	sketch of an Aharonov-Bohm interferometer whose purpose is to
	measure the spatial coherence length $\xi$ of the charged particle.
	We will see that its coherence length decreases as the coupling
	between the particle and the environment increases, even at
	zero temperature.

\subsection{Description of the device}
	Consider a perfect two-dimensional waveguide.
	In the longitudinal direction ($z$-axis) the particle
	is free, while in the transverse direction ($q$-axis)
	it is confined by a harmonic potential (see fig.~\ref{fig:guide}). 
	Hence the hamiltonian of the particle in the waveguide is~:
		\begin{equation}%
		\op{H}_{\mathrm{guide}}
		= 
		\underbrace{
		\frac{\op{p}^2}{2m}
		+
		\frac{m\Omega^2 }{2}\op{q}^2
		}_{\op H_{\mathrm{transv.}}}
		+
		\frac{\op{p}_z^2}{2m}
		\end{equation}%
	We furthermore consider that at $z=0$ are connected two 1-dimensional leads
	separated by a distance $2x$, as  shown in fig.~\ref{fig:guide}.
	A particle moving in the waveguide can enter
	the ``upper'' one or the ``lower'' one. In lead 1 (2)
	the particle is described as one-dimensional,
	moving along the $z_1$-axis ($z_2$-axis)~:
		\begin{equation}%
		\op{H}_{\mathrm{lead}_i}
		= 
		\frac{\op{p}_{z_i}^2}{2m}
		%
		\end{equation}%
	The Hilbert
	space of the particle is the direct sum of the space of the 
	waveguide and the space of the leads \cite{albeverio}.
		\begin{equation}%
		\mathcal{H}_{\mathrm{tot}}
		=
		\mathcal{H}_{\mathrm{guide}}
		\oplus
		\mathcal{H}_{\mathrm{lead1}}
		\oplus
		\mathcal{H}_{\mathrm{lead2}}
		\end{equation}%
	Thus, a quantum state of the particle
	is given by writing the state of the particle in these
	three spaces~:
		\begin{equation}%
		\left(
		\begin{array}{c}
			\ket{\psi}_{\mathrm{guide}} \\
			\ket{\psi}_{\mathrm{lead1}} \\
			\ket{\psi}_{\mathrm{lead2}}
			\end{array}
			\right)
		\end{equation}%
	\label{sec:coupl}
	We suppose that the coupling between each lead
	and the waveguide corresponds to a ``tunnel'' coupling.
	The wire and the waveguide interact only in a very
	localised region.
	The total hamiltonian (coupling included) 
	can be written as~:
		\begin{eqnarray*}
		\op{H}_{\mathrm{tot}}	
		&=&
		\left(
		\begin{array}{ccc}
			\op{H}_{\mathrm{guide}}
			&
			\op V_1
			&
			\op V_2
			\\
			\op V_1^+
			&
			\op{H}_{\mathrm{lead1}}
			&
			0
			\\
			\op V_2^+
			&
			0
			&
			\op{H}_{\mathrm{lead1}}
			\end{array}
		\right)
		\\
		\op V_1 &=& \alpha \ket{v_1} \bra{f_1}
		\\
		\op V_2 &=& \alpha \ket{v_2} \bra{f_2}
		\end{eqnarray*}
	The coupling energy is represented by the off-diagonal terms.
	The states $\ket{f_1}$ and $\ket{f_2}$ of the lead are supposed
	to be very localised near the leads' origin. The 
	states in the waveguide $\ket{v_1}$ and $\ket{v_2}$ are
	localised near the  contact points $z=0,q=x$ and $z=0,q=-x$.
	The real constant $\alpha$ represents  the strength of
	the coupling. When $\alpha$ vanishes, the tunnelling between
	the waveguide and the leads disappears.
	By putting a variable flux $\Phi$ between
	the two leads, one can induce a phase shift $\phi$ between them.
	The probability to detect the particle as a function
	of the flux oscillates by varying the flux (see fig.~\ref{fig:guide}).
	This Aharonov-Bohm interferometer \cite{sakurai}
	is nearly equivalent to the Young double slit.
	
\subsection {Transverse coupling with the environment}

	Now, let us suppose that there is a region
	in the waveguide where the particle is
	\emph{transversely} coupled with an external environment
	(for instance, through a capacitor for a charged particle),
	see fig.~\ref{fig:guide}. In the case of a linear resistor
	and a perfect capacitor, it is straightforward
	(but tedious) to show that the spectral function
	$\mu(\omega)$ of the environment is~:
		\begin{equation}%
		\mu(\omega) =
		\frac{2e^2}{\pi\omega^2 l^2 C}
		\times
		\frac{1/RC}{\omega^2 + (1/RC)^2}
		\end{equation}%
	where $e$ is the charge of the particle, $l$ is
	the distance between the plates of the capacitor
	$C$ is the capacitance and $R$ the resistance of the
	linear resistor. For large $R$, the environment
	has typical strength $\eta=Re^2/l^2$ and cut-off
	frequency $\Omega_c=1/RC$. Finally we should note
	that the transversal frequency of the waveguide
	is renormalised by the capacitive coupling as 
	follows~:
		\begin{equation}%
		m\Omega^2 = m\Omega^2_\mathrm{guide}
		- \frac{e^2}{l^2C}
		\end{equation}%
	where $\Omega_\mathrm{guide}$ is the frequency
	of the guide without coupling.	
	
	The environment acting in the $q$-direction 
	is represented by a linear resistor,
	and is well described by the model considered
	in sec.~\ref{sec:matrice}. 
	We suppose that the dissipative region is large 
	enough in the $z$-direction so that any particle 
	entering on the left will have reached thermal 
	equilibrium with the bath in the $q$-direction
	before leaving it on the right. It means that if 
	a particle enters the region with a pure state
	$\ket{\psi}\otimes\ket{k}$ where $\ket{\psi}$
	is any transverse state and $\ket{k}$ is
	a plane wave in the $z$-direction, it will leave
	the dissipative region in a state
	described by a density operator~:
		\begin{equation}%
		\op \sigma \otimes \ket{k}\bra{k}
		\end{equation}%
	where $\op{\sigma}$ is the particle state in the  $q$-direction,
	given by eq.~\ref{eq:sigma}. We also suppose that
	the distance between the end of the dissipative region
	and the contact points of the pair of leads is much smaller
	than $\hbar k / m\Omega$.

	More precisely, the total hamiltonian in the guide is~:
		\begin{eqnarray}
		\nonumber
		H(q,p,\varphi,\pi) &=& 
		\frac{p_z^2}{2m} +
		\frac{p^2}{2m} +
		\frac{m\Omega^2}{2}q^2
		\\
		\nonumber
		&+&
		\sum_{i=1}^{N} 
		\left\{
		\frac{\pi_i^2}{2\mu_i} +
		\frac{\mu_i \omega_i^2}{2}
		\left(
		\varphi_i - \lambda(z) q
		\right)^2
		\right\}
		\end{eqnarray}
	where  $\lambda(z)=1$ if the particle is inside the dissipative 
	region and $\lambda(z)=0$ otherwise. Except on the boundaries of
	the dissipative region, we suppose that the coupling between particle 
	and environment is transverse. Thus, except on the boundaries, 
	the transverse and the longitudinal dynamics are independent.
	Scattering on the boundaries is possible. On the one
	hand, $\lambda(z)$
	has no singularities, thus the dependence on $z$ of the
	stationary states of the system is continuous and the state
	of the particle is the same on both sides of any boundary.
	On the other hand, one can show that we can neglect 
	scattering by assuming that the initial longitudinal energy fulfils
	the following two conditions~:
		\begin{eqnarray*}
		\frac{p_z^2}{2m} &\gg&
			\av{q^2}\sum\mu_i\omega_i^2
			\\
		\frac{p_z^2}{2m} &\gg&
			\left(
			\av{q^2}\sum\mu_i\omega_i^2\av{\epsilon_i}
			\right)^{1/2}
		\end{eqnarray*}			
	where $\av{\epsilon_i}$ is the mean energy
	of the $i$-th oscillator of the environment. With these
	assumptions, longitudinal and transverse degrees of freedom  are
	independent along the whole guide. In this case the $z$-coordinate,
	the longitudinal dynamics is trivial, it simply behaves as translation in
	time.

	Experimentally, the initial pure  state
	$\ket{\psi}\otimes\ket{k}$ can be prepared by putting a
	particle with fixed wave-vector $k$ in a single-channelled
	waveguide. At this stage, the only possible transverse state
	is the ground state. However, if the number of transverse
	channels grows \emph{adiabatically} along the $z$-axis,  it
	turns out that the particle will necessarily remain in the
	ground state \cite{glazman}. In this case $\ket{\psi} = \ket{0}$, the
	transverse ground state of the waveguide.
	
\subsection{Interference fringes}

	In appendix~\ref{sec:annexe2}, it is shown that in the limit
	when the incident wave vector satisfies the condition that its
	longitudinal energy is much larger than its transverse energy,
	the probability that the particle be found  in the cross-point
	of the two leads, after a measure is in term of any transverse 
	state~$\op\sigma$~:
		\begin{eqnarray*}
		P(\phi) &=&
			|\tau|^2 \times 
			\Big(
			\sigma(x,x) + \sigma(-x,-x)
			\\
			&&
			\sigma(-x,x) e^{i\phi} +
			\sigma(x,-x) e^{-i\phi}
			\Big)
		\end{eqnarray*}
	where $\sigma(x,x')=\av{x|\op\sigma|x'}$ is the position
	representation of $\op\sigma$. Similarly, let $P_{1}$ and $P_{2}$ be the probability
	to find the particle respectively in lead~1 and
	lead~2~:
		\begin{eqnarray*}
		P_{1} &=& |\tau|^2 \sigma(x, x) \\
		P_{2} &=& |\tau|^2 \sigma(-x, -x)
		\end{eqnarray*}
	We define the normalised contrast $C$ of the interference fringes
	by~:
		\begin{equation}
		C^2 = \frac{\av{P^2}-\av{P}^2}{2P_{1} P_{2}}, 
		\quad \textrm{where }
		\av{f} = 
			\frac{1}{2\pi}
			\int_{0}^{2\pi}
			f(\phi)\dd \phi
		\end{equation}
	By using the value of $\sigma(q,q')$, discussed in
	appendix \ref{sec:annexe1}~:
		\begin{displaymath}
		\sigma(x,x') =
		\frac{1}{2\pi\av{\op q^2}}
		e^{ -\frac{1}{2} 
			\left[
			\frac{\av{\op p^2}}{\hbar^2} (x-x')^2
			+
			\frac{1}{4\av{\op q^2}}(x+x')^2
			\right]}
		\end{displaymath}
	we obtain~:
		\begin{eqnarray*}
		P(\phi) &=& 2|\tau|^2 \sigma(x,x) \times
			\left(
			1 + e^{-{x^2}/{2\xi^2}} \cos \phi 
			\right)
		\\
		&&\textrm{where\quad}
		\xi^2 = 
			\frac{\hbar}{2\tilde m\Omega} \times
			\frac{1}{1 - \tanh\frac{\hbar\Omega}{2k\tilde T}}
		\end{eqnarray*}
	Finally, we obtain for the normalised contrast~:
		\begin{equation}%
		C
		= \exp{-\frac{x^2}{2\xi^2}}
		\end{equation}%
	This expression defines the transverse \emph{coherence length}
	$\xi$ of the particle. It shows that the particle
	is capable to interfere along a maximum length
	of $\xi$.
		
	%
	For the environment discussed in sec.~\ref{sec:matrice}, consider the environment discussed in 
	section~\ref{sec:exple1}. We can explicitly calculate $\xi$ for
	that particular environment (see appendix~\ref{sec:annexe2}). 
	Fig.~\ref{fig:leff} shows the transverse coherence length of
	the particle, as function of the (real) temperature, for
	various values of the coupling  strength $\eta$. In the
	zero-temperature limit, saturation occurs and $\xi(T=0)$ has a finite value, showing
	that long range interferences are not possible. At $T=0$, 
	we find the following limit expression
	of $\xi$, for small $\eta$~:
		\begin{equation}%
		\xi^2(0) \sim
			\frac	{\hbar}{\eta}
			\times
			\frac	{\pi}
				{4\ln{(\Omega_c/\Omega})}
		\end{equation}%
	Even at zero temperature, the interference fringes are
	destroyed because of the coupling between particle and 
	environment. The behaviour of the particle is similar to that
	of a particle at strictly positive temperature.	 This effect
	becomes more pronounced as the coupling strength increases.
	
	Finally, let us remark that the evolution in the wave-guide is
	not reversible because the environment has infinite 
	number of degrees of freedom. This should be 
	taken into account when transport phenomenas
	are considered.

\section{Conclusion}

        We considered a simple exactly solvable model
	of a particle coupled to the environment. Our purpose
	was to discuss the influence of the environment on the particle
	at thermal equilibrium. After obtaining the reduced density
	operator of the particle, we proposed a sketch for a simple device
	capable to measure the spatial coherence length of the particle
	through Aharonov--Bohm interference measurement. We saw that the
	coupling between the particle and the environment destroys long
	range interferences and reduces its spatial coherence length.
	Even in the zero temperature limit, this effect remains and
	grows with the coupling between the particle and the environment.

        We finally discuss two possible experimental realisations of our model.

        Let us first consider a confined
	two-dimensional electron gas, where the coherence length of the
	electron is about several hundred nanometres,
	comparable to the typical size of a small mesoscopic device \cite{datta}.
	In view of this scenario, an external circuit giving rise to electrodynamic noise
	could play the role of the environment.
	The main problem is the construction of the point
	contact between the leads and the waveguide. Actually, if the
	contact is very localised, one may loose a significant part of
	the measured signal. Furthermore, a direct interpretation of such an experiment could be masked by
	many-body effects: the Pauli principle has to be taken into account when
	treating decoherence in a two-dimensional electron gas, along the lines presented in~\cite{marquardt}.

        Another example is a cold atomic gas, localised in a magnetic
	trap. The atomic gas is a quantum system, that looses
	coherence in the presence of fluctuations of the trapping
	potential, induced by fluctuations of the applied magnetic
	field (the "environment"). Interference-like experiments, similar to the ones discussed here,
	have been proposed to study these decoherence
	phenomena~\cite{schroll}. The environment in~\cite{schroll} is a high-temperature one;
	it would be interesting to extend the discussion to the case of a low-temperature environment, such that
	the effects discussed in the present paper become important.

\section{Acknowledgements}
	We thank M.~Büttiker, L.~Lévy and V.~Renard for valuable discussions.
	Our research was sponsored by Institut Universitaire de France
	and CNRS--ATIP.

\appendix
\section{Reduced density operator of the particle}	
\label{sec:annexe1}

\subsection{Classical modes}

	The classical system defined by eq.~\ref{eq:hamtot} is made of $N+1$
	harmonic oscillators coupled one to each other,  but since the
	hamiltonian is quadratic and positively defined, 
	one can decompose the system as a set of $N+1$
	\emph{independent} harmonic oscillators (the eigen-modes of the
	total system).

\subsubsection{Matrix notations}
	Let us first rewrite the classical hamiltonian \ref{eq:hamtot}
	in a matrix form, where
	we separate the $N+1$ positions from the $N+1$ momenta~: 
		\begin{eqnarray*}
		H 
		&=&
		\frac{1}{2}(P|P) + \frac{1}{2}(Q|A|Q)
		\end{eqnarray*}
	where~:
		\begin{eqnarray*}
		|P) &=& \mat{c}{
			\frac{p}{\sqrt{m}} \\ 
			\vdots \\
			\frac{\pi_i}{\sqrt{\mu_i}} \\ 
			\vdots
			}
			,\quad
		|Q) = 	\mat{c}{
			q\sqrt{m}  \\ 
			\vdots \\
			\phi_i \sqrt{\mu_i} \\ 
			\vdots
			}
			\\
		A &=&	\mat{cccc} {
				\Omega^2 + \sum \omega_i^2\frac{\mu_i}{m}
					&
					\dots 
					&
					-\omega_i^2\sqrt{\frac{\mu_i}{m}}
					&
					\dots
				\\
				\vdots
				&
				\ddots
				&
				&
				\\
				-\omega_i^2\sqrt{\frac{\mu_i}{m}}
					&
					&
					\omega_i^2
					&
					0
				\\
				\vdots
					&
					0
					&
					&
					\ddots
				}
		\end{eqnarray*}
	We note $|0),\dots,|N)$ the canonical base in
	$\mathbf{R}^{N+1}$, so that the vectors $|Q)$ and $|P)$
	in $\mathbf{R}^{N+1}$ have components~:
		\begin{equation}
		\begin{displaystyle}
		\begin{array}{rcl@{\quad\quad\quad}rcl}
		 q	\sqrt{m}&=& (0|Q)
		&
		 \pi_i /\sqrt{\mu_i}&=& (i|P)
		\\
		 \varphi_i \sqrt{\mu_i} &=& (i|Q)
		&
		 p /\sqrt{m}&=& (0|P)
		\\
		\end{array}
		\end{displaystyle}
		\end{equation}
	The eigen-vectors $|U_j)$ and eigen-values $\nu_j^2$ of 
	the symmetric and positive matrix $A$ verify~:
		\begin{eqnarray}
		A|U_j) &=& \nu_j^2 |U_j) \\
		I &=& \sum_{j=0}^{N} |U_j)(U_j|
		\end{eqnarray}
	We can, now, rewrite the hamiltonian with respect
	to a new set of ``normal'' positions and momenta~:
		\begin{equation}
		\label{eq:hamtot-bis}
		H 
		= 
		\sum_{j=0}^{N}  H_j 
		=
		\frac{1}{2}
		\sum_{j=0}^{N}
		\left\{
		 y_j^2 
		+
		\nu_j^2  x_j^2
		\right\}
		\end{equation}
	with the new canonical coordinates
	$ x_j = (U_j|Q)$ and $ y_j = (U_j|P)$. 
	Conversely, because $s_{ij}=(i|U_j)$
	is an orthogonal matrix~:
		\begin{equation}
		\begin{array}{rcl@{\quad}rcl}
		 q \,\sqrt{m}		&=& \displaystyle{\sum_{j=0}^{N} (0|U_j)  x_j}
		&
		 p / \sqrt{m}		&=& \displaystyle{\sum_{j=0}^{N} (0|U_j)  y_j}
		\\
		 \varphi_i \,\sqrt{\mu_i}	&=& \displaystyle{\sum_{j=0}^{N} (i|U_j)  x_j}
		&
		 \pi_i / \sqrt{\mu_i}	&=& \displaystyle{\sum_{j=0}^{N} (i|U_j)  y_j}
		\end{array}
		\end{equation}


\subsubsection{Reduced resolvent}
	In sec.~\ref{sec:q2p2}, eq.~\ref{eq:defq2} and
	\ref{eq:defp2}
	we need to compute~:
		\begin{equation}%
		M_f = \sum_{j=0}^N (0|U_j)^2 f(\nu_j^2)
		\end{equation}%
	where $f$ is any smooth function.
	We will follow a standard calculation,
	using the resolvent, to explicitly obtain that
	expression (see \cite{exner,fano,cohen}). Consider the resolvent $R(z)$
	of the matrix $A$, defined by as $R(z)=(z-A)^{-1}$, with 
	$z\in\mathbf{C}$. In the eigen-base of $A$, $R(z)$ can be
	written as~:
		\begin{equation}%
		R(z) = \sum_{j=0}^N \frac{|U_j)(U_j|}{z-\nu_j^2}
		\end{equation}%
	For $z$ near the real axis, 
	we can set $z=\epsilon+i\kappa$ and take the limit
	$\kappa\rightarrow0$ in the imaginary part of $R$~:
		\begin{equation}%
		\im 
		\left\{ 
		\lim_{\kappa\rightarrow0^+}R(\epsilon+i\kappa)
		\right\}		 
		= 
		-\pi \sum_{j=0}^N |U_j)(U_j| \delta(\epsilon - \nu_j^2)
		\end{equation}%
	Since $\nu_j^2>0$, only the positive $\epsilon$ are concerned
	and we can take $\epsilon=u^2$, we obtain~:
		\begin{equation}
		M_f
		=
		- \frac{1}{\pi}
		\int_0^{\infty}
		\left\{ \lim_{\kappa\rightarrow0^+} 
		\im (0|R(u^2+i\kappa)|0)
		\right\}
		f(u^2)  
		\dd u^2
		\label{eq:defmoy}
		\end{equation}
	Let us decompose the matrix $A$ as the sum of
	its diagonal and its off-diagonal part~: $A=A_0+V$. One
	can write $R$ as~:
		\begin{equation}%
		R = R_0 + R_0 V R \textrm{, where } R_0(z) = (z-A_0)^{-1}
		\end{equation}%
	Consider~:
		\begin{eqnarray}
		\nonumber
		(0|R|0) 
		&=& (0|R_0|0) + (0|R_0|0)(0|VR|0)
		\\
		\nonumber
		&=& (0|R_0|0)
			\left[ 
			1 + \sum_{i=1}^N 
			(0|V|i)(i|R|0)
			\right]
		\\
		&=& \frac{1}{z-\Omega^2-\sum_{i=1}^N \omega_i^2\mu_i/m}
			\\
			\nonumber
			&& \quad 
			\times
			\left[ 
			1 - \sum_{i=1}^N 
			\omega_i^2\sqrt{\frac{\mu_i}{m}}(i|R|0)
			\right]
		\\
		\nonumber
		\\
		\nonumber
		(i|R|0) 
		&=& (i|R_0|0) + (i|R_0|i)(i|V|0)(0|R|0)
		\\
		&=& 	- 
			\frac{1}{z-\omega_i^2} \, 
			\omega_i^2\sqrt{\frac{\mu_i}{m}} \,
			(0|R|0)
		\end{eqnarray}	
	Finally, by combining the last two expressions, we obtain~:
		\begin{equation}%
		(0|R(z)|0)^{-1} = 
			z - 
			\Omega^2 -
			z\sum_{i=1}^N \frac{\omega_i^2 \mu_i/m}{z -\omega_i^2}
		\end{equation}%

\subsubsection{Continuum limit}

	In that form, the expression of $(0|R|0)$ is exact but nearly 
	unusable. However, remember that we are mainly interested in the
	case where the number of oscillators in the 
	environment is very large, so we can take the continuum
	limit $N\rightarrow\infty$ and replace the set
	of frequencies $\omega_i$ with a continuous variable $\omega$.
	The mass distributions is described by the smooth function
	$\mu(\omega)$ defined such that $\mu(\omega)\dd\omega$ is
	the mass of the oscillators with frequency 
	between $\omega$ and $\omega+\dd\omega$. In that limit, 
		\begin{equation}%
		(0|R(z)|0)^{-1} = 
			z - 
			\Omega^2
			-
			z\int_{0}^{\infty} 
				\frac{\omega^2 \mu/m}{z-\omega^2} 
				\dd\omega
		\end{equation}%
	Now, we can take $z=u^2+i\kappa$ and 
	consider the limit $\kappa\rightarrow0$~:
		\begin{displaymath}
		\lim_{\kappa\rightarrow0^+}
		(0|R(u^2+i\kappa)|0) =
		\frac{1}{
			u^2-\Omega^2-\Delta(u)
			+iu\,\Gamma(u)
			}
		\end{displaymath}
	where~:
		\begin{eqnarray*}
			\Delta(u) &=& \displaystyle{
				\mathcal{P}\int_0^{\infty}
				\frac{\omega^2 u^2/m}{u^2-\omega^2}
				\mu(\omega)\dd \omega
				}
				\\
			\Gamma(u) &=& \displaystyle{
				\frac{\pi u^2}{2m}
				}\mu(u)
		\end{eqnarray*}
	Finally, by putting the last expression in
	eq.~\ref{eq:defmoy}, we obtain in the continuum limit~:
		\begin{equation}
		\label{eq:useful}
		\sum_{j=0}^N (0|U_j)^2 f(\nu_j^2)
		\longrightarrow
		\frac{2}{\pi}
		\int_0^{\infty}
			\frac{\Gamma u^2  f(u^2) \dd u}
			{(u^2-\Omega^2-\Delta)^2+\Gamma^2u^2}
		\end{equation}

\subsection{Fluctuation of the position and the momentum of the particle}
	\label{sec:q2p2}	
	We consider now the quantum version of the system defined by
	eq.~\ref{eq:hamtot}, \emph{i.e.} a quantum oscillator coupled
	to a quantum environment. We replace the classical variables
	with operators; in the case when the whole system is in equilibrium
	at temperature $T$, its density operator is the following~:
		\begin{equation}%
		\op{\rho} = \frac{1}{Z}e^{-\op{H}/kT}
		\textrm{\quad where \quad}
		Z = \tr \op \rho
		\end{equation}%
	From eq.~\ref{eq:hamtot-bis}
	we can write $\op \rho$ as~:
		\begin{eqnarray}
		\op \rho &=& 
		\frac{e^{-\op{H}_0/kT}}{Z_0}
		\otimes
		\frac{e^{-\op{H}_1/kT}}{Z_1}
		\otimes
		\cdots
		\otimes
		\frac{e^{-\op{H}_N/kT}}{Z_N}
		\\
		&&
		\textrm{where \ }
		Z_j = \tr \left( e^{-\op{H}_j/kT} \right)
		\end{eqnarray}
	The mean square fluctuations of the position of
	the particle in the state $\op\rho$ are~:
		\begin{eqnarray}
		\nonumber
		\av{\op{q}^2}_T 
		&=& 
		\frac{1}{m}
		\av{ 
			\sum_{j=0}^{N} (0|U_j) \op x_j
			\sum_{j'=0}^{N} (0|U_{j'}) \op x_{j'}
			}
		\\
		\nonumber
		&=&
			\frac{1}{m}
			\sum_{j=0}^{N} (0|U_j)^2 \av{\op x_j^2}
		\\
		\label{eq:defq2}
		&=&
			\sum_{j=0}^{N}  (0|U_j)^2
				\left\{ 
				\frac{\hbar}{2 m \nu_j}\coth\frac{\hbar\nu_j}{2kT}
				\right\}
		\end{eqnarray}
	Where in the last step we have replaced $\av{\op x_j^2}$
	with its value, obtained for a simple 1-dimensional oscillator.	
	In the same way, we can find
	the mean square fluctuations of the position of
	the particle~:
		\begin{eqnarray}
		\label{eq:defp2}
		\av{\op{p}^2}_T 
		&=& 
			\sum_{j=0}^{N}  (0|U_j)^2
				\left\{ 
				\frac{m\hbar\nu_j}{2}\coth\frac{\hbar\nu_j}{2kT}
				\right\}
		\end{eqnarray}
	We see that these expressions have the form of
	eq.~\ref{eq:useful},  so, for a continuous environment, we can
	explicitly compute $\av{\op q^2}$ and $\av{\op p^2}$  for any
	mass distribution $\mu(\omega)$.
	These results have already been obtained (see
	\cite{weiss,buttiker}),
	following a different approach, namely 
	the fluctuation-dissipation theorem.

\subsection {Reduced density operator of the particle}

	The reduced density operator $\op{\sigma}$ of the particle 
	is defined
	as the trace over the environment of $\op\rho$~:
		\begin{equation}%
		\op\sigma = \tr_{\mathrm{env.}}{\op\rho}
		\end{equation}%
	Since the total hamiltonian is quadratic
	in position and momentum, $\op{\rho}$ is Gaussian in these
	operators; when we
	trace over the degrees of freedom of the environment, 
	$\op{\sigma}$ remains Gaussian and we can write~:
		\begin{equation}%
		\op{\sigma} = e^{
			a\op{p}^2 +
			b\op{q}^2 +
			c
			}
		\end{equation}%
	with unknown coefficients $a$, $b$, $c$. 
	Note that there is not a term of the form $\op p\op q$
	because the total hamiltonian is invariant by time reversal
	and therefore $\op \rho$ and $\op \sigma$ are also invariant.
	We can rewrite $\op{\sigma}$ in a canonical form
	with unknown coefficients $\tilde T$ and $\tilde m$~:
		\begin{equation}%
		\op{\sigma} = \frac{1}{\tilde Z} \exp \left\{
			- \frac{1}{k\tilde T}
				\left(
				\frac{\op{p}^2}{2\tilde m} + 
				\frac{\tilde m \Omega^2}{2}\op{q}^2
				\right)
				\right\}
		\end{equation}%
	This is not a standard definition of $\tilde T$, $\tilde m$ and
	$\op H_{\mathrm{eff}}$. Since only two of the coefficients
	$\tilde T$, $\tilde m$ and $\Omega$ are independent, we have
	chosen to avoid having an effective frequency  $\tilde \Omega$.
	With this definition the effective  hamiltonian has the same
	spectrum as the hamiltonian of
	the particle. Also, keeping $\Omega$ the same
	in both hamiltonians is the only way to
	ensure that there is a linear canonical transformation
	of particle hamiltonian leading to $\op H_{\mathrm{eff}}$.	
	With this requirement, parameters
	$\tilde T$ and $\tilde m$ are uniquely defined. 
	
	Let us now determine $\tilde Z$, $\tilde T$ and $\tilde m$.
	$\tilde Z$ is a normalisation constant that ensures that
	$\tr \op\sigma=1$. We can explicitly calculate~:
		\begin{eqnarray}
		&&
		\av{\op p^2} 
			\displaystyle{\mathop{=}^{\mathrm{def}}} 
			\tr (\op p^2\op \sigma) 
			= \frac{\hbar\tilde m \Omega}{2}\coth{\frac{\hbar\Omega}{2k\tilde T}}
		\\
		&&
		\av{\op q^2} 
			\displaystyle{\mathop{=}^{\mathrm{def}}}
			\tr (\op q^2\op \sigma) 
			= \frac{\hbar}{2\tilde m \Omega}\coth{\frac{\hbar\Omega}{2k\tilde T}}
		\end{eqnarray}
	Conversely, we obtain~:
		\begin{eqnarray}
		\label{eq:tildeT}
		k\tilde T &=& 
			\frac{\hbar\Omega}{2}
			\frac{1}{\Argth{\sqrt{\frac{\hbar^2}{4\av{\op p^2}\av{\op q^2}}}}}
		\\
		\label{eq:tildem}
		\tilde m &=& 
			\sqrt{\frac{\av{\op p^2}}{\Omega^2\av{\op q^2}}}
		\end{eqnarray}
	Finally, since according to eq.~\ref{eq:defq2},
	\ref{eq:defp2} and
	\ref{eq:useful}, $\av{\op q^2}$
	and $\av{\op p^2}$ can be written as function of the
	mass distribution of the environment $\mu(\omega)$, we 
	can do either for $\tilde m$ and $\tilde T$. Finally,
	we have obtained the expression of the reduced density operator,
	which, in position representation, can be written as~:
		\begin{equation}
		\sigma(q,q') =
		\frac{1}{2\pi\av{\op q^2}}
		e^{ -\frac{1}{2} 
			\left[
			\frac{\av{\op p^2}}{\hbar^2} (q-q')^2
			+
			\frac{1}{4\av{\op q^2}}(q+q')^2
			\right]}
		\label{eq:sigma-pos}
		\end{equation}

	Let us remark that the evolution in the wave-guide is
	not reversible because the environment has infinite 
	number of degrees of freedom. This should be 
	taken into account when transport phenomenas
	are considered.

\section{Interference fringes} \label{sec:annexe2}
\subsection{Scattering coefficients}

	Consider the coupling between the waveguide and the leads, 
	discussed in sec.~\ref{sec:coupl}. We will
	rewrite it in the following way~:
		\begin{eqnarray*}
		\op V_1 &=& \alpha (\ket{g_1}\otimes\ket{f}) \bra{f_1}
		\\
		\op V_2 &=& \alpha (\ket{g_2}\otimes\ket{f}) \bra{f_2}
		\end{eqnarray*}
	The states $\ket{f_1}$ and $\ket{f_2}$ of the lead are supposed
	to be very localised near the leads' origin. The transverse
	states in the waveguide $\ket{g_1}$ and $\ket{g_2}$ are
	localised near the attach points $q=x$ and $q=-x$ and the
	longitudinal state $\ket{f}$ is localised around the origin
	$z=0$. 
	The real constant $\alpha$ represents  the strength of
	the coupling. Note that the later states
	are normalised and not Dirac peaks, they are such that~:
		\begin{eqnarray*}
		\av{f|f} &=& 1 \\
		\av{z|f} &\sim& \sqrt{\varepsilon} \ \delta(z) 
		\end{eqnarray*}
	where $\epsilon$ is the typical spatial width of the state $\ket{f}$.
	We are interested by the stationary state of the particle,
	made of an incident wave, a reflected wave and three transmitted
	waves (one through the waveguide and two through the 
	pair of leads). Suppose that the incident wave is
	in the $n'$-th channel ($n'$-th excited transverse state). It may
	be transmitted and reflected in the other channels, 
	we can write the stationary wave function in the
	$n$-th channel and the leads as~:
		\begin{equation}%
		\psi_n(q,z) = \chi_n(q)\phi_n(z) 
		\end{equation}%
	where $\chi_n(q)$ is the stationary wave-function of the
	transverse ``harmonic oscillator'' with energy $ E_n =
	\hbar\Omega(n + 1/2)$; the longitudinal part $\phi_n(z)$ of the
	wave-function   is made by an incident, a reflected and  a
	transmitted wave~:
		\begin{eqnarray*}
		&&\phi_n(z) =
			\left\{
			\begin{array}{l@{}c@{\ }l}
			\textrm{if $z<0$\quad} & &
			\delta_{nn'}e^{ik_nz} 
			+r_{n}e^{-ik_nz}
			\\
			\textrm{if $z>0$\quad} & &
			t_{n}e^{ik_nz}
			\end{array}
			\right.
		\end{eqnarray*}
	$\delta_{nn'}$ is the Kronecker symbol, $t_n$ and $r_n$
	respectively are the transmission and the reflection
	coefficients in the $n$-th channel and $k_n$ is
	the corresponding wave-vector~:
		\begin{equation}%
		k_n = \sqrt{k^2 + 2m( E_{n'}- E_{n})/\hbar^2}
		\end{equation}%
	Finally, 
	the wave-function in each lead is just a transmitted
	plain-wave, it can be written as~:
		\begin{eqnarray*}
		\sigma_1(z_1) = s_1 e^{i\kappa|z_1|} \\
		\sigma_2(z_2) = s_2 e^{i\kappa|z_2|}
		\end{eqnarray*}
	where $s_1$, $s_2$
	are the transmission coefficients in the guide and
	$\kappa$ the corresponding wave-vector~:
		\begin{equation}%
		\kappa = \sqrt{k^2 + 2m E_{n'}/\hbar^2}
		\end{equation}%
	Since this 
	state is supposed to be stationary, it is an eigen-vector
	of the hamiltonian in \ref{sec:coupl}, we can write~:
		\begin{eqnarray*}
		-\frac{\hbar^2}{2m}\sigma_1'' +
			\alpha\varepsilon^{3/2}\delta(z_1)\sum_{n=0}^{\infty}\chi_n(x)\phi_n(0)
			&=&  E\sigma_1
		\\
		-\frac{\hbar^2}{2m}\sigma_2'' +
			\alpha\varepsilon^{3/2}\delta(z_2)\sum_{n=0}^{\infty}\chi_n(-x)\phi_n(0)
			&=&  E\sigma_2
		\\
		 E_n\phi_n - \frac{\hbar^2}{2m}\phi_n'' 
			+ \alpha\varepsilon^{3/2}\delta(z)\sigma_1(0)\bar\chi_n(x)\delta(z_1) 
			&& \\
			+ \alpha\varepsilon^{3/2}\delta(z) \sigma_2(0)\bar\chi_n(-x)\delta(z_2) 
			&=&  E\phi_n
		\end{eqnarray*}
	These three equations, and the continuity condition
	on $\phi_n$, leads to the following set of linear equations~:
		\begin{equation}%
		\begin{array}{rcl}
		-\hbar^2ik_n (t_n - \delta_{nn'} + r_n)/2m
			+\alpha \varepsilon^{3/2} s_2 \chi_n(-x)
			&=& 0
		\\
		-\hbar^2i\kappa s_1 /2m
			+ \alpha \varepsilon^{3/2} \sum_{n=0}^{\infty}
			\chi_n(x)t_n 
			&=& 0
		\\
		-\hbar^2i\kappa s_2 /2m
			+ \alpha \varepsilon^{3/2} \sum_{n=0}^{\infty}
			\chi_n(-x)t_n 
			&=& 0
		\\
		\delta_{nn'} + r_n - t_n &=& 0
		\end{array}
		\end{equation}%
	finally, we obtain~:
		\begin{eqnarray*}
		s_1 &=&
			\frac{\alpha \varepsilon^{3/2} m}{i\hbar^2\kappa}
			\frac{1}{R^2 + |Z|^2}
			[ R\chi_{n'}(x) - Z \chi_{n'}(-x) ]
		\\
		s_2 &=&
			\frac{\alpha \varepsilon^{3/2} m}{i\hbar^2\kappa}
			\frac{1}{R^2 + |Z|^2}
			[ R\chi_{n'}(-x) - \bar{Z} \chi_{n'}(x) ]
		\end{eqnarray*}
	where~:
		\begin{eqnarray}
		R &=& 
			1 + \frac{\alpha^2\varepsilon^3m^2}{\hbar^2\kappa^2}
			\sum \frac{1}{\lambda_{n}} |\chi_n(x)|^2
		\\
		Z &=& 
			\frac{\alpha^2\varepsilon^3m^2}{\hbar^2\kappa^2}
			\sum \frac{1}{\lambda_{n}}\chi_n(x)\bar\chi_n(-x)
		\\
		\lambda_{n} &=& \sqrt{1-\frac{2m E_n}{\hbar^2/\kappa^2}}
		\end{eqnarray}
	Consider, now, the much simpler case where
	$\hbar^2k^2\gg E_{n'}$. In that case,
	$\lambda_n\rightarrow{}1$, and thus
	$Z\rightarrow{}0$
	and $R\rightarrow 1+(\alpha\epsilon m / \hbar k)^2$,
	finally, the transmission coefficients 
	can be written as~:
		\begin{eqnarray}
		s_1 &=& \tau \ \chi_{n'}(x) 
				\\
		s_2 &=& \tau \ \chi_{n'}(-x)
		\end{eqnarray}
	where~:
		\begin{equation}%
		\tau = 
			\frac{\alpha\epsilon m}{i\hbar k}
			\frac{\epsilon^{1/2}}{1 + 
				\Big(\displaystyle{
					\frac{\alpha\epsilon m}{\hbar k}}
					\Big)^2}		
		\end{equation}%

\subsection{Interference fringes}

	For an incident state in the waveguide of the form
	$\ket{\chi_n}\otimes\ket{k}$, where $\ket{\chi_n}$
	is an eigenstate of $\op H_{\mathrm{transv.}}$
	and $\ket{k}$ a plane wave in the $z$-direction,
	one associates the transmitted states in the lead,
	written~:
		\begin{equation}%
		\left(
			\begin{array} {c}
			s_{1n} \ket{k_1} \\
			s_{2n} \ket{k_2} 
			\end{array}
			\right)
		\end{equation}%
	where $\ket{k_1}$, $\ket{k_2}$ are plane waves
	and $(s_{1n},s_{2n}) \in \mathbf{C}^2$ are
	transmission coefficients which may depend
	on $n$ and $k$. 
	Thus, by the separation principle, if the particle in the guide
	is described by a density operator~:
		\begin{eqnarray*}
		\op\sigma \otimes \ket{k}\bra{k} &=&
		\left( \sum_{n,n'} \sigma_{nn'} \ket{\chi_n}\bra{\chi_{n'}}
		\right)
		\otimes 
		\ket{k}\bra{k}
		\end{eqnarray*}
	the transmitted state is described by the density operator~:
		\begin{eqnarray*}
		\op w = |\tau|^2
		\left(
			\begin{array} {rr}
			\sigma(x,x)   \, \ket{k_1}\bra{k_1} &
			\sigma(x,-x)  \, \ket{k_1}\bra{k_2} 
			\\
			\sigma(-x,x)  \, \ket{k_2}\bra{k_1} &
			\sigma(-x,-x) \, \ket{k_2}\bra{k_2} 
			\end{array}
			\right)
		\end{eqnarray*}
	The probability that the particle be found 
	in the cross-point of the two leads,
	after a measure is~:
		\begin{eqnarray}
		I(\phi) 
		&=& 
			\bra{z_1} 
			\bra{z_2} 
			\op w
			\ket{z_1} 
			\ket{z_2}
			\\
		&=& |\tau|^2
			\Big[ 
			\sigma(x,x) + \sigma(-x,-x)
			\\
		&&+
			\sigma(-x,x) e^{i\phi} +
			\sigma(x,-x) e^{-i\phi}
			\Big]
		\end{eqnarray}
	By replacing $\sigma(x,x')$ with its value form 
	eq.~\ref{eq:sigma-pos} we can explicitly
	evaluate $I(\phi)$~:
		\begin{eqnarray*}
		I(\phi) &=& 2|\tau|^2\sigma(x,x) \times
			\left(
			1 + e^{-{x^2}/{2\xi^2}} \cos \phi 
			\right)
		\\
		&&\textrm{where\quad}
		\xi^2 = 
			\frac	{\av{\op q^2}\hbar^2}
				{4\av{\op p^2}\av{\op q^2} - \hbar^2}
		\end{eqnarray*}
	By using eq.~\ref{eq:defq2} and eq.~\ref{eq:defp2} we see
	that $\xi$ is related to the effective temperature
	and to the effective mass in the following way~:
		\begin{equation}%
		\xi^2 = 
			\frac{\hbar}{2\tilde m\Omega} \times
			\frac{1}{1 - \tanh\frac{\hbar\Omega}{2k\tilde T}}
		\end{equation}%
	Since, in appendix~\ref{sec:annexe1}, 
	$\tilde T$ and $\tilde m$ were written as functions
	of the internal parameters of the environment, we can
	do either with $\xi$.

\end{document}